\documentclass[12pt]{article}
\usepackage{graphicx}

\newcommand{\Frac}[2]{\frac{\displaystyle #1}{\displaystyle #2}}
\newcommand{\beq}{\begin{equation}}
\newcommand{\eeq}{\end{equation}}
\newcommand{\beqn}{\begin{eqnarray}}
\newcommand{\eeqn}{\end{eqnarray}}
\newcommand{\beqns}{\begin{eqnarray*}}
\newcommand{\eeqns}{\end{eqnarray*}}

\begin{document}
\begin{titlepage}
\begin{center}

\hfill USTC-ICTS-03-14\\
\hfill  November 2003

\vspace{2.5cm}

{\Large {\bf   Long-distance contribution to the forward-backward
asymmetry in  decays $K^+\to\pi^+ \ell^+\ell^-$
\\}} \vspace*{1.0cm}
 {  Dao-Neng Gao$^\dagger$} \vspace*{0.3cm} \\
{\it Interdisciplinary Center for Theoretical Study and Department
of Modern Physics, University of Science and Technology of China,
Hefei, Anhui 230026 China}

\vspace*{1cm}
\end{center}
\begin{abstract}
\noindent The long-distance contribution via the two-photon
intermediate state to the forward-backward asymmetries in decays
$K^+\to\pi^+\ell^+\ell^-$ ($\ell=e$ and $\mu$) has been studied
within the standard model. In order to evaluate the dispersive
part of the $K^+\to\pi^+\gamma^*\gamma^*\to\pi^+\ell^+\ell^-$
amplitude, we employ a phenomenological form factor to soften the
ultraviolet behavior of the transition. It is found that, this
long-distance transition, although subject to some theoretical
uncertainties, can lead to significant contributions to the
forward-backward asymmetries, which could be tested in the future
high-precise experiments.
\end{abstract}

\vfill
\noindent

$^{\dagger}$ E-mail:~gaodn@ustc.edu.cn
\end{titlepage}
Rare kaon decays provide interesting information on the structure
of the weak interactions at low energies \cite{BK00, BB01, DI98}.
Among them, the flavor-changing neutral-current processes
$K^\pm\to\pi^\pm \ell ^+ \ell^-$ ($\ell=e$ and $\mu$), induced at
the one-loop level in the standard model (SM), are well suited
both to explore the quantum structure of the SM and to search for
new physics beyond it \cite{DEIN95}. The total decay rates for
these transitions are dominated by the long-distance contribution
via one-photon exchange, which have been successfully described
within the framework of chiral perturbation theory \cite{EP95} up
to $O(p^6)$ in terms of a vector form factor \cite{DEIP98} fixed
by experiments \cite{E86599}.

Many useful observables in these transitions, such as P- and
T-violating muon polarization effects in $K^+\to\pi^+\mu^+\mu^-$
\cite{SW90, ANBG91, BB94, DV99, DG01, DM01}, as well as the
CP-violating charge asymmetries in $K^\pm\to\pi^\pm \ell^+\ell^-$
\cite{Messina02, DG02}, were investigated in the past literatures.
Recently another interesting observable, the forward-backward
asymmetry $A_{\rm FB}$ in $K^+\to\pi^+\ell^+\ell^-$, has been
studied \cite{CQ03}. As pointed out by Chen, Geng, and Ho
\cite{CQ03}, present experimental constraints allow large values
of $A_{\rm FB}$'s; thus it is expected that the measurement of
these asymmetries in future experiments could be very interesting
both to test the SM and to probe new physics scenarios. The
purpose of this paper is devoted to the analysis of this
forward-backward asymmetry in the SM, which is induced by the
long-distance transition via the two-photon intermediate state,
$K^+\to\pi^+\gamma^*\gamma^*\to\pi^+\ell^+\ell^-$. Since at
present the dispersive part of the
$K^+\to\pi^+\gamma^*\gamma^*\to\pi^+\ell^+\ell^-$ amplitude, which
contains the logarithmic divergences, cannot be evaluated in a
model-independent way, we will employ a phenomenological form
factor proposed in Refs. \cite{DIP98,BDI03} to soften the
ultraviolet behavior of this transition.  As we shall see, this
long-distance transition will lead to a scalar form factor and an
extra antisymmetric vector form factor under the exchange of the
lepton momenta ($p_+\leftrightarrow p_-$),  and both of them  can
induce contributions to the forward-backward asymmetries in
$K^+\to\pi^+\ell^+\ell^-$ decays.

The general invariant amplitude for
$K^+(p)\to\pi^+(p_\pi)\ell^+(p_+)\ell^-(p_-)$ can be parameterized
as \cite{ANBG91, CQ03} \beq\label{IA1} {\cal
M}=F_S\bar{\ell}\ell+i
F_P\bar{\ell}\gamma_5\ell+F_Vp^\mu\bar{\ell}\gamma_\mu\ell+
F_Ap^\mu\bar{\ell}\gamma_\mu\gamma_5\ell,\eeq where $p$, $p_\pi$,
$p_\pm$ are the four-momenta of $K^+$, $\pi^+$, and $\ell^\pm$,
and $F_S$, $F_P$, $F_V$, and $F_A$ are scalar, pseudoscalar,
vector, and axial-vector form factors, respectively. The
differential decay rate takes the form \beqn\label{rate1}
\frac{d\Gamma}{dz d\cos\theta}=\frac{m_K^5\beta_\ell
\lambda^{1/2}(1,z,r_\pi^2)}{2^8\pi^3}\left\{\left|\frac{F_S}{m_K}\right|^2
z \beta_\ell^2+\left|\frac{F_P}{m_K}\right|^2
z+|F_V|^2\frac{1}{4}\lambda(1,z,r_\pi^2)(1-\beta_\ell^2
\cos^2\theta)\right.\nonumber \\
\left.
+|F_A|^2\left[\frac{1}{4}\lambda(1,z,r_\pi^2)(1-\beta^2_\ell
\cos^2\theta)+4 r_\ell^2\right]\right.\nonumber\\\left.+\frac{{\rm
Re}(F_S F_V^*)}{m_K}
2r_\ell~\beta_\ell\lambda^{1/2}(1,z,r_\pi^2)\cos\theta +\frac{{\rm
Im}(F_PF_A^*)}{m_K}2r_\ell (r_\pi^2-1-z) \right\},
 \eeqn
 where $\lambda(a,b,c)=a^2+b^2+c^2-2(ab+ac+bc)$,
 $r_\ell=m_\ell/m_K$, $r_\pi=m_\pi/m_K$, $z=(p_++p_-)^2/m_K^2$, $\beta_\ell=\sqrt{1-4r_\ell^2/z}$,
 and $\theta$ is the angle between the three
 momentum of the kaon and the three momentum of $\ell^-$ in the
 dilepton rest frame.
The phase space in terms of  $z$ and $\cos\theta$ is given by
\beq\label{phasespace}
 4r_\ell^2\le
 z\le(1-r_\pi)^2,\;\;\;\;\;-1\le \cos\theta\le 1.\eeq
 Thus the forward-backward asymmetry is defined as
 \beq\label{AFB1}
 A_{\rm FB}(z)=\Frac{\int^1_0 \left(\frac{d\Gamma}{dzd\cos\theta}\right)d\cos\theta-\int^0_{-1}
 \left(\frac{d\Gamma}{dzd\cos\theta}\right)d\cos\theta}{\int^1_0 \left(\frac{d\Gamma}
 {dzd\cos\theta}\right)d\cos\theta+\int^0_{-1}
 \left(\frac{d\Gamma}{dzd\cos\theta}\right)d\cos\theta}.\eeq

As seen from Eqs. (\ref{rate1}) and (\ref{AFB1}), in general only
two form factors $F_S$ and $F_V$ will play relevant roles in
obtaining the significant $A_{\rm FB}$. In the SM, one-photon
exchange transition $K^+\to\pi^+ \gamma^*\to \pi^+\ell^+\ell^-$
dominates the form factor $F_V$; while two-photon exchange
transition $K^+\to\pi^+\gamma^*\gamma^*\to\pi^+\ell^+\ell^-$ can
contribute to both $F_S$ and $F_V$. It has been shown in
\cite{DEIP98}, one-photon exchange contribution to $F_V$ can be
written as \beq\label{FV1gamma}F_V^{\gamma}=-\frac{\alpha
G_F}{2\pi}(a_++b_+z)- \frac{\alpha}{2\pi m_K^2}W^{\pi\pi}_+,\eeq
where the real parameters $a_+$ and $b_+$ encode local
contributions starting from $O(p^4)$ to $O(p^6)$ in chiral
perturbation theory, and the experimental measurement of
$K^+\to\pi^+ e^+ e^-$ by BNL E865 \cite{E86599} has determined
them to be
\beq\label{ABvaule}a_+=-0.587\pm0.010,\;\;\;\;b_+=-0.655\pm0.044.\eeq
The non-analytic term $W^{\pi\pi}_+$ denotes pion-loop
contribution, which is estimated to $O(p^6)$ using the physical
$K^+\to\pi^+\pi^+\pi^-$ data, and its full expression can be found
in Ref. \cite{DEIP98}.

The general invariant amplitude for $K^+\to\pi^+\gamma\gamma$ is
given by \beq\label{IA2gamma1}{\cal
A}[K^+(p)\to\pi^+(p_\pi)\gamma(q_1,\epsilon_1)\gamma(q_2,
\epsilon_2)]=\epsilon_{1\mu}(q_1)\epsilon_{2\nu}(q_2){\cal
M}^{\mu\nu}(p,q_1,q_2),\eeq and ${\cal M}^{\mu\nu}$ can be
decomposed as \beqn\label{M2gamma1} {\cal
M}^{\mu\nu}&=&\frac{A(y,z)}{m_K^2}(q_2^\mu q_1^\nu-q_1\cdot q_2
g^{\mu\nu})+\frac{2B(y,z)}{m_K^4}(p\cdot q_1 q_2^\mu p^\nu+p\cdot
q_2 p^\mu q_1^\nu -p\cdot q_1 p\cdot q_2
g^{\mu\nu}\nonumber\\&&-q_1\cdot q_2 p^\mu
p^\nu)+\frac{C(y,z)}{m_K^2}\varepsilon^{\mu\nu\alpha\beta}q_{1\alpha}q_{2\beta}
+\frac{D(y,z)}{m_K^4}[\varepsilon^{\mu\nu\alpha\beta}(p\cdot q_2
q_{1\alpha}+p\cdot q_1q_{2\alpha})p_{\beta}\nonumber\\
&&+(p^\mu\varepsilon^{\nu\alpha\beta\sigma}+p^\nu\varepsilon^{\mu\alpha\beta\sigma})p_\alpha
q_{1\beta}q_{2\sigma}],\eeqn where $y=p\cdot(q_1-q_2)/m_K^2$ and $
z=(q_1+q_2)^2/m_K^2$. For our purposes, now we are concerned about
$A(y,z)$ and $B(y,z)$ amplitudes since $C(y,z)$ and $D(y,z)$ are
irrelevant to the present discussion. Within the framework of
chiral perturbation theory, the amplitude $A(y,z)$ will receive
non-vanishing contribution at $O(p^4)$, which has been computed in
Ref. \cite{EPR88}; while the leading order contribution to
$B(y,z)$ starts from $O(p^6)$, and only the unitarity corrections
and vector resonance contributions to it were evaluated
\cite{DP96, CDM93}. It is easy to find that, via the transition
$K^+\to\pi^+\gamma^*\gamma^*\to\pi^+\ell^+\ell^-$, the leading
order $A$ amplitude only contributes to $F_S$, and the $B$
amplitude contributes to both $F_S$ and $F_V$, which is similar to
the case of $K_L\to\pi^0\gamma^*\gamma^*\to\pi^0\ell^+\ell^-$
\cite{EPR88, DHV87, FR89HS93, DG95}. Therefore one can expect that
the contribution of the scalar form factor $F_S$ in
$K^+\to\pi^+\ell^+\ell^-$ due to the two-photon intermediate state
is dominantly given by the $O(p^4)$ $A$ amplitude, and in the
following we will neglect high order contributions to $F_S$ from
the $B$ and $O(p^6)$ $A$ amplitudes.

The leading $\Delta I=1/2$ $O(p^4)$ $A(y,z)$ amplitude for
$K^+\to\pi^+\gamma\gamma$ can be expressed as \cite{EPR88}
 \beq\label{amp1p4} A(y,z)=\frac{G_8m_K^2
\alpha}{2\pi z}\left[(r_\pi^2-1-z)F\left({z\over r_\pi^2
}\right)+(1-r_\pi^2-z)F(z)+\hat{c} z\right],
 \eeq where $|G_8|=9.2\times 10^{-6}$ GeV$^{-2}$.
$F(z/r_\pi^2)$ and $F(z)$ are generated from $\pi$ and $K$ loop
diagrams respectively, which could be defined as
\beq\label{loop}F(x)=\left\{\begin{array}{cc}
1-\Frac{4}{x}\arcsin^2\left(\Frac{\sqrt{x}}{2}\right) & x\leq
4,\\\\1+\Frac{1}{x}\left(\ln{\Frac{1-\sqrt{1-4/x}}{1+\sqrt{1-4/x}}}+i\pi\right)^2&x\geq
4.\end{array}\right. \eeq $\hat{c}$ in eq. (\ref{amp1p4}) is from
$O(p^4)$ non-anomalous local counter-terms \cite{KMW90}, which is
a quantity $O(1)$.   The first observation of the decay
$K^+\to\pi^+\gamma\gamma$ was reported in \cite{BNLE787}, and a
maximum likelihood fit of $\hat{c}$ to the decay spectrum using
the absolutely normalized rate has been performed to fix the value
of $\hat{c}$: without the unitarity corrections, $\hat{c}=1.6\pm
0.6$ with $\chi^2/{\rm DOF}=6.3/7$; with the unitarity
corrections,
 $\hat{c}=1.8\pm0.6$ with $\chi^2/{\rm DOF}=4.6/7$.

In Ref. \cite{DP96}, the $O(p^6)$ contribution to
$K^+\to\pi^+\gamma\gamma$ including unitarity corrections from
$K^+\to\pi^+\pi^+\pi^-$ and local terms generated by vector
resonance exchange has been evaluated. As pointed out by
D'Ambrosio and Portol\'es \cite{DP96}, the unitarity corrections
are relevant while the vector meson contributions are likely to be
negligible. Thus the corresponding $B$ amplitude can be written as
\cite{DP96} \beqn\label{Bamplitude}
B(y,z)=\frac{\alpha}{\pi}\left\{\frac{1}{3
r_\pi^4}(4\zeta_1+\xi_1)\left[-\frac{1}{6}\left(1+2
\ln\frac{m_\pi^2}{\mu^2}\right)+\frac{z}{18r_\pi^2}-\frac{2r_\pi^2}{z}
~F\left(\frac{z}{r_\pi^2}\right)\right.\right.\nonumber\\
\left.\left.+\frac{1}{3}\left(\frac{z}{r_\pi^2}-10\right)R\left(\frac{z}{r_\pi^2}\right)\right]\right\},
\eeqn where parameters $\zeta_1$ and $\xi_1$ have been determined
from the phenomenology of $K^+\to\pi^+\pi^+\pi^-$ \cite{KMW91},
the mass scale $\mu$ is generally taken as $m_\rho$ in the
numerical calculation, and
\beq\label{loop2}R(x)=\left\{\begin{array}{cc}
-\Frac{1}{6}+\Frac{2}{x}-\Frac{2}{x}\sqrt{4/x-1}\arcsin\left(\Frac{\sqrt{x}}{2}\right)
& x\leq 4,\\\\-\Frac{1}{6}+\Frac{2}{x}+\Frac{\sqrt{1-4/x}}{x}
\left(\ln{\Frac{1-\sqrt{1-4/x}}{1+\sqrt{1-4/x}}}+i\pi\right)&x\geq
4.\end{array}\right. \eeq

Now following the similar procedure as in the case of the neutral
channels $K_L\to\pi^0\gamma^*\gamma^*\to\pi^0\ell^+\ell^-$
\cite{DG95, BDI03}, one can get the amplitudes from the $A$ and
$B$ terms via the two-photon intermediate state for the charged
channels, which read \beqn\label{A2gammatoee}{\cal
M}(K^+\to\pi^+\gamma^*\gamma^*\to\pi^+\ell^+\ell^-)^A=\frac{ie^2}{m_K^2}\int\frac{d^4q}{(2\pi)^4}
\frac{A(y,z)}{q^2(Q-q)^2[(p_--q)^2-m_\ell^2)]}\nonumber\\
\times~ \bar{u}(p_-)\{[3 q^2-2 (Q+p_-)\cdot q+s]\gamma_\mu q^\mu-2
m_\ell~ Q\cdot q+2 m_\ell~ q^2\}v(p_+), \eeqn and
\beqn\label{B2gammatoee} {\cal
M}(K^+\to\pi^+\gamma^*\gamma^*\to\pi^+\ell^+\ell^-)^B=\frac{2ie^2}{m_K^4}\int\frac{d^4q}{(2\pi)^4}
\frac{B(y,z)\bar{u}(p_-)\gamma_\mu
v(p_+)}{q^2(Q-q)^2[(p_--q)^2-m_\ell^2)]}\nonumber\\
\times\left\{p^\mu[2p\cdot q p_-\cdot Q-2p\cdot p_-q\cdot
Q-p\cdot(p_+-p_-)q^2]\right.\nonumber\\
\left. +q^\mu[2p\cdot q p\cdot(p_+-p_-)-2p\cdot Qp\cdot q+2p\cdot
q p\cdot q ]\right\},
 \eeqn
 where $Q=p_++p_-$, $s=Q^2$, and $q$ is the loop momentum for the internal photon. Since
 the integrals in Eqs. (\ref{A2gammatoee}) and (\ref{B2gammatoee}) are logarithmically
 divergent, only their absorptive part
contributions can be calculated unambiguously. Actually for the
off-shell photons, the $A(y,z)$ and $B(y,z)$
 amplitudes corresponding to the on-shell photons,  should be replaced
by $A[y,z, q^2, (Q-q)^2]$ and $B[y,z, q^2, (Q-q)^2]$,
respectively. At present, there is no model-independent way to
obtain these off-shell form factors.  Analogous to the analysis of
$K_L\to\pi^0\gamma^*\gamma^*\to\pi^0 e^+ e^-$ presented in Ref.
\cite{BDI03}, we employ the following ansatz to regularize the
above integrals\beqn\label{ansatz} A[y,z, q^2,
(Q-q)^2]=A(y, z)\times f[q^2, (Q-q)^2], \nonumber\\ \\
B[y,z, q^2, (Q-q)^2]=B(y, z)\times f[q^2,(Q-q)^2]\nonumber\eeqn
with the form factor \beq\label{fqQ-q} f[q^2, (Q-q)^2]=1+a
\left[\frac{q^2}{q^2-m_V^2}+\frac{(Q-q)^2}{(Q-q)^2-m_V^2}\right]+
b\frac{q^2(Q-q)^2}{(q^2-m_V^2)[(Q-q)^2-m_V^2]}\eeq is defined in
analogy with the analysis of the
$K_L\to\gamma^*\gamma^*\to\mu^+\mu^-$ in Ref. \cite{DIP98}, and
the parameters $a$ and $b$ are expected to be $O(1)$ by naive
dimensional chiral power counting. This structure is dictated by
the assumption that vector meson dominance (VMD) plays a crucial
role in the matching between short and long distance physics (in
the numerical calculation $m_V$ is conventionally chosen to be the
$\rho$ mass, i.e., $m_V\simeq 770$ MeV). As shown in \cite{BDI03},
in order to obtain the ultraviolet convergent integrals, we need
to impose the condition\footnote{when one includes the $O(p^6)$
contribution to the $A$ amplitude, this condition will be not
enough to guarantee the convergent integral in Eq.
(\ref{A2gammatoee}); however, as discussed above and as a good
approximation, we neglect this high order contribution.}
\beq\label{condition} 1+2a+b=0. \eeq In a special case for
$a=-b=-1$, the form factor (\ref{fqQ-q}) will be identical to the
one adopted in Ref. \cite{DG95} for $K_L\to\pi^0\gamma^*\gamma^*$.
It is then straightforward to perform the integrals in Eqs.
(\ref{A2gammatoee}) and (\ref{B2gammatoee}) after including the
form factors in Eq. (\ref{ansatz}). Neglecting terms which are
suppressed by powers of $1/m_V^2$ and eliminating $b$ by means of
Eq. (\ref{condition}), we get \beq\label{A2gammatoee2}{\cal
M}(K^+\to\pi^+\gamma^*\gamma^*\to\pi^+\ell^+\ell^-)^A=F_S^{\gamma\gamma}
\bar{u}(p_-)v(p_+), \eeq and \beqn\label{B2gammatoee2}{\cal
M}(K^+\to\pi^+\gamma^*\gamma^*\to\pi^+\ell^+\ell^-)^B=F_V^{\gamma\gamma}p^\mu
\bar{u}(p_-)\gamma_\mu v(p_+), \eeqn where
\beqn\label{FS2gamma}F_S^{\gamma\gamma}=\frac{\alpha m_\ell
A}{4\pi m_K^2}\left\{\frac{7}{2}-6a-3\ln{\frac{r_V^2}{z}}
 -\int^1_0 dx\int^x_0
dy\left(\frac{(1+\frac{2r_\ell^2}{z})(1-x)^2}{\frac{r_\ell^2}{z}
(1-x)^2-y(x-y)}\right.\right.\nonumber\\
\left.\left.-(12-9x)\ln\left[\frac{r_\ell^2}{z}(1-x)^2-y(x-y)\right]\right)\right\}\eeqn
is the scalar form factor, \beqn\label{FV2gamma}
F_V^{\gamma\gamma}&=&\frac{\alpha B}{2\pi
m_K^4}p\cdot(p_+-p_-)\left\{\frac{2}{3}\ln\frac{r_V^2}{z}-\frac{8}{9}
+\frac{4}{3}a\right.\nonumber\\&&\left.-\int^1_0 dx\int^x_0 dy~
(2-x) \ln\left[\frac{r_\ell^2}{z}(1-x)^2-y(x-y)\right] \right\}
\eeqn is the extra vector form factor [relative to the one-photon
exchange contribution to $F_V$ in Eq. (\ref{FV1gamma})] via
two-photon intermediate state,  and $r_V=m_V/m_K$.
 The absorptive parts of $F_S^{\gamma\gamma}$ and $F_V^{\gamma\gamma}$ for
on-shell two photons can be extracted directly from Eqs.
(\ref{FS2gamma}) and (\ref{FV2gamma}) as \beq\label{absorptiveFS}
F_S^{\gamma\gamma}|_{\rm absorptive}=\frac{i\alpha m_\ell A}{4\pi
m_K^2}\frac{1}{\beta_\ell}\ln\frac{1-
\beta_\ell}{1+\beta_\ell},\eeq and
\beq\label{absorptiveFV}F_V^{\gamma\gamma}|_{\rm
absorptive}=\frac{i\alpha B}{8
m_K^4}p\cdot(p_+-p_-)\frac{1}{\beta_\ell^2}\left[\frac{2}{3}+\frac{2}{\beta_\ell^2}-
\left(\frac{1}{\beta_\ell^2}-\beta_\ell^2\right)\frac{1}{\beta_\ell}\ln\frac{1+
\beta_\ell}{1-\beta_\ell}\right].\eeq Actually these results [Eqs.
(\ref{absorptiveFS}) and (\ref{absorptiveFV})] are model
independent, which are consistent with ones using the methods
presented in Refs. \cite{EPR88, DHV87, FR89HS93}.

Contributions to the forward-backward asymmetry $A_{\rm FB}$ due
to the interference between $F_S^{\gamma\gamma}$ and $F_V^\gamma$
can be easily derived from Eqs. (\ref{rate1}) and (\ref{AFB1}) as
\beq\label{AFB2} A^a_{\rm FB}(z)=\frac{m_K^4r_\ell~\beta_\ell^2
\lambda(1,z,r_\pi^2)}{2^7\pi^3} {\rm Re}(F_S^* F_V^\gamma
)/(d\Gamma/dz), \eeq where $F_S^*$ denotes
$(F_S^{\gamma\gamma})^*$, and $d\Gamma/dz$ is the differential
decay rate after integrating the angle $\theta$ in Eq.
(\ref{rate1}) (neglecting the contributions from $F_P$ and $F_A$).
Meanwhile, since $F_V^{\gamma\gamma}$ is  proportional to $p\cdot
(p_+ - p_-)$, the significant asymmetry $A_{\rm FB}$ can also be
generated from the interference between $F_V^{\gamma\gamma}$ and
$F_V^\gamma$. Using the relation \beq\label{pkp-p} p\cdot
(p_+-p_-)=-\frac{m_K^2}{2}\beta_\ell \lambda^{1/2}(1,z,r_\pi^2)
\cos\theta, \eeq one can get the asymmetry \beq\label{AFB3}
A^b_{\rm FB}(z)=\frac{m_K^7 \beta_\ell^2\lambda^2(1,z,r_\pi^2)(1-
\beta_\ell^2/2))}{2^{10} \pi^3}{\rm
Re}(\tilde{f}_V^*F_V^\gamma)/(d\Gamma/dz),\eeq where
$\tilde{f}_V=- F_V^{\gamma\gamma}/p\cdot(p_+-p_-)$.\vspace{0.5cm}

One can find that there is a free parameter $a$ in the expressions
of $F_S^{\gamma\gamma}$ and $F_V^{\gamma\gamma}$ [Eqs.
(\ref{FS2gamma}) and (\ref{FV2gamma})], which should be $O(1)$
from the naive dimensional chiral power counting, however, cannot
be fixed from both the theoretical and phenomenological analysis
at present. It is expected that the future experimental study of
$K^+\to\pi^+\gamma\gamma^*\to\pi^+\gamma\ell^+\ell^-$ could
provide some interesting information on it \cite{FG99}. In the
following, we will take $a=-1, ~0, +1$, respectively, to
illustrate the numerical results for $A^a_{\rm FB}(z)$ and
$A^b_{\rm FB}(z)$. Numerical calculations show that contributions
from both $F_S^{\gamma\gamma}$ and $F_V^{\gamma\gamma}$ to the
decay rate of $K^+\to\pi^+\ell^+\ell^-$ are negligible, consistent
with one-photon exchange dominant mechanism in this decay, and the
values of $F_S^{\gamma\gamma}$ from Eq. (\ref{FS2gamma}) are
smaller than the experimental bound on the scalar form factor,
$|F_S/(G_F m_K)|\le 6.6\times 10^{-5}$, for $K^+\to\pi^+ e^+ e^-$
given in \cite{E86599}.

Since $A^a_{\rm FB}$ in Eq. (\ref{AFB2}) is proportional to
$m_\ell^2$, the scalar contribution to the forward-backward
asymmetry in $K^+\to\pi^+ e^+ e^-$ is strongly suppressed. The
numerical analysis gives that it is at most $O(10^{-4})$. However,
$A^a_{\rm FB}$ in $K^+\to\pi^+\mu^+\mu^-$ can be $O(10^{-2})$,
which has been plotted in Fig. 1. Interestingly in the region of
large $z$, the sign of $A_{\rm FB}^a$ for the muon mode is
sensitive to the value of $a$. This is not surprising because, in
Eq. (\ref{AFB2}), Re$(F_S^* F^\gamma_V)$ in this region can change
sign for the value of $a$ varying from $1$ to $-1$, as shown in
Fig. 2, while the differential decay rate $d\Gamma/dz$ in this
region, dominated by $F_V^\gamma$, almost remains the same for
different $a$, as shown in Fig. 3. Thus the measurement of
$A^a_{\rm FB}$ in $K^+\to\pi^+\mu^+\mu^-$ in the region of large
$z$ might impose some interesting constraints on the value of $a$.

The forward-backward asymmetries $A_{\rm FB}^b$'s in
$K^+\to\pi^+e^+ e^-$ and $K^+\to\pi^+ \mu^+\mu^-$ have been
plotted in Fig. 4 and Fig. 5, respectively. Now there is no
similar $m_\ell^2$ suppression in $A^b_{\rm FB}$ as that in
$A_{\rm FB}^a$. It is seen that  $A^b_{\rm FB}$ can be
$O(10^{-3})$ for the electron mode, and about from $10^{-4}$ to
$10^{-3}$ for the muon mode. On the other hand, if we only
consider the absorptive parts of $F_S^{\gamma\gamma}$ [Eq.
(\ref{absorptiveFS})] and $F_V^{\gamma\gamma}$ [Eq.
(\ref{absorptiveFV})] in Eqs. (\ref{AFB2}) and (\ref{AFB3}),
$A^a_{\rm FB}$ for the muon mode can also be $O(10^{-2})$ in the
region of large $z$ but with the positive sign; $A^b_{\rm FB}$'s
for both the electron and muon modes will be very small, which are
only $O(10^{-5})$.

\begin{figure}[t]
\begin{center}
\includegraphics[width=14cm,height=15cm]{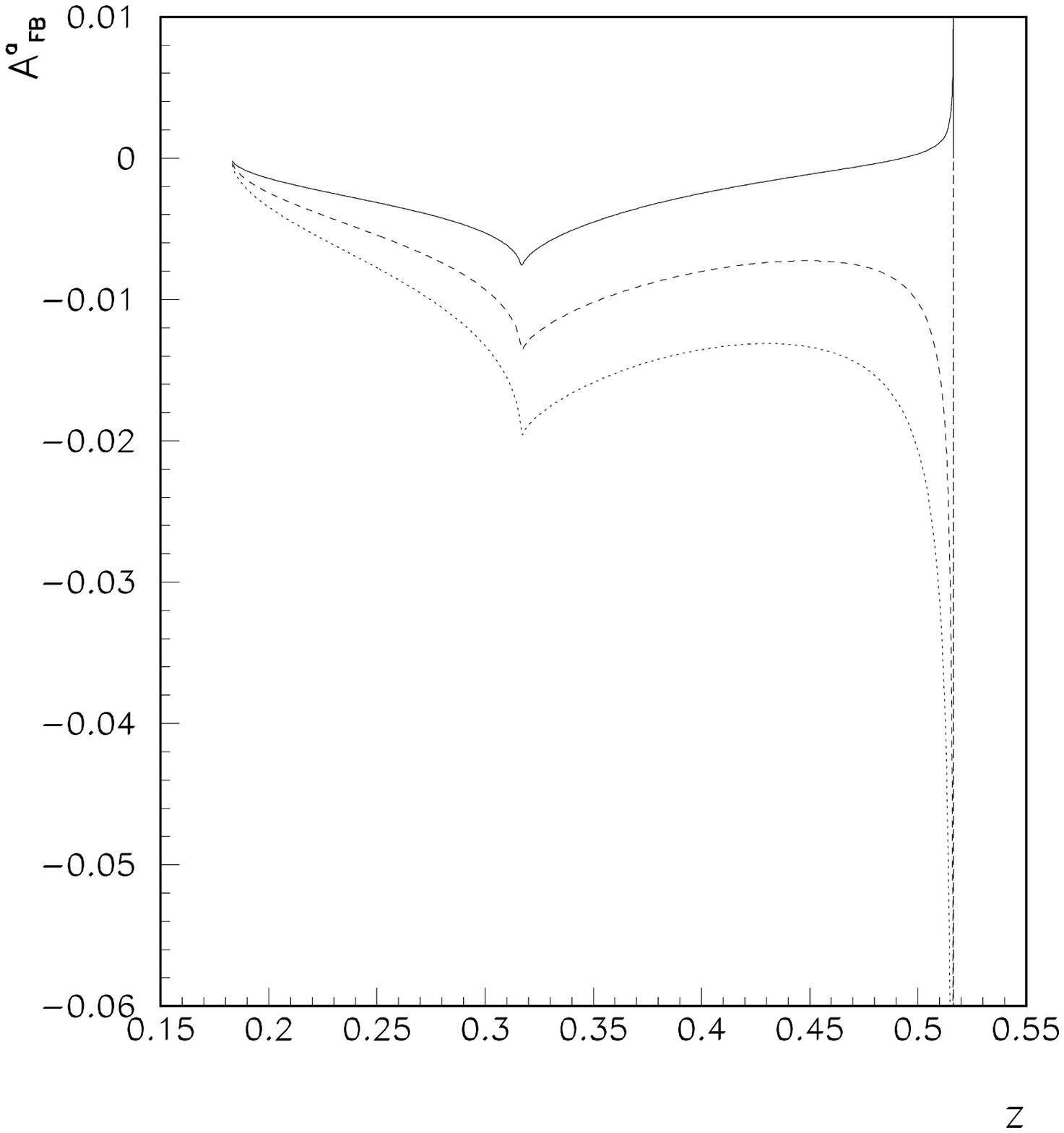}
\end{center}
\caption{Forward-backward asymmetry $A^a_{\rm FB}$ in decay
$K^+\to\pi^+\mu^+\mu^-$ as a function of $z$. The full line is for
$a=-1$, the dashed line for $a=0$, and the dotted line for $a=1$.
}
\end{figure}

\begin{figure}[t]
\begin{center}
\includegraphics[width=14cm,height=15cm]{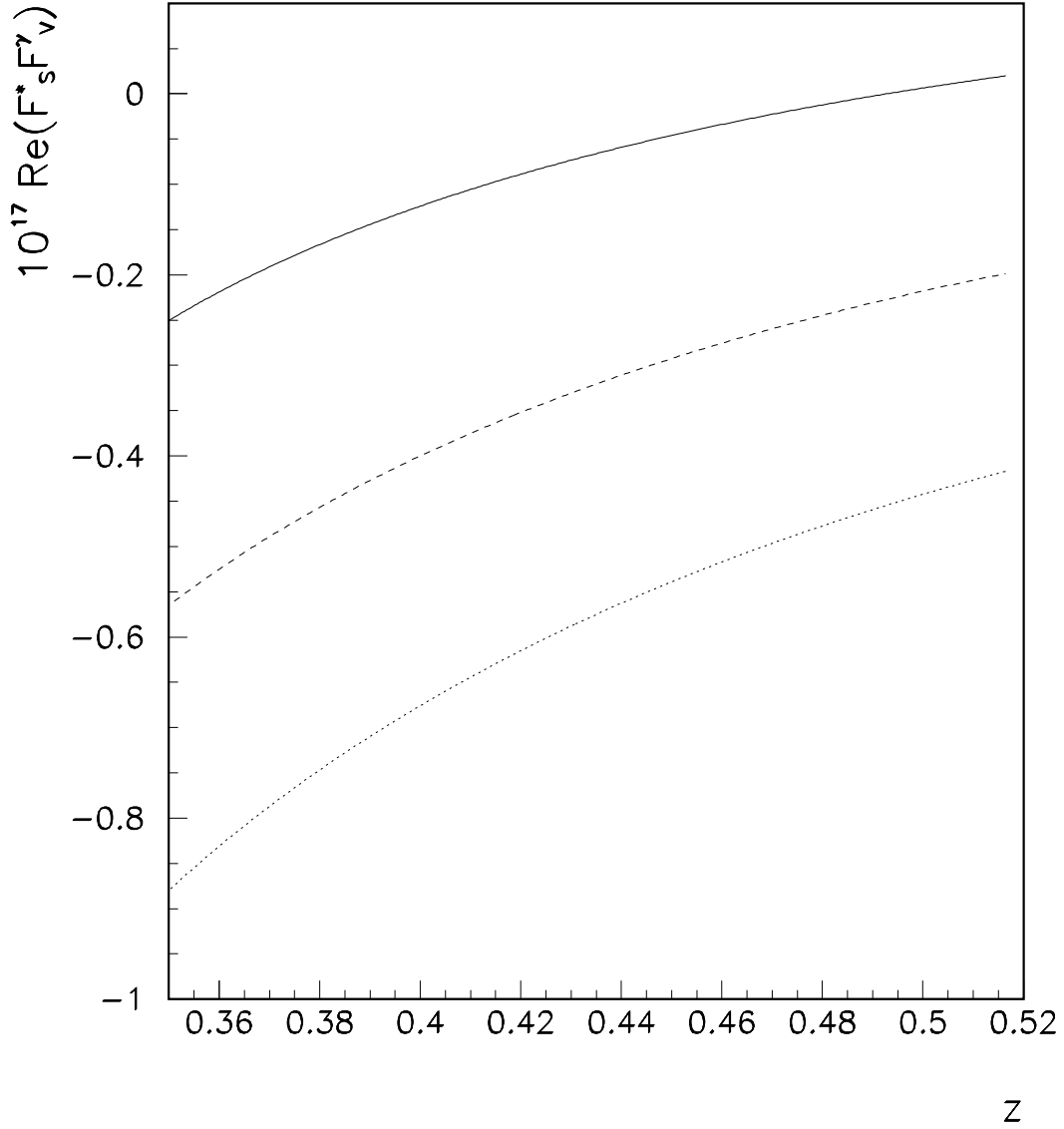}
\end{center}
\caption{${\rm Re}(F_S^* F_V^\gamma)$ in decay
$K^+\to\pi^+\mu^+\mu^-$ as a function of $z$ with $0.35\le z\le
(1-r_\pi)^2$. The full line is for $a=-1$, the dashed line for
$a=0$, and the dotted line for $a=1$. }
\end{figure}

\begin{figure}[t]
\begin{center}
\includegraphics[width=14cm,height=15cm]{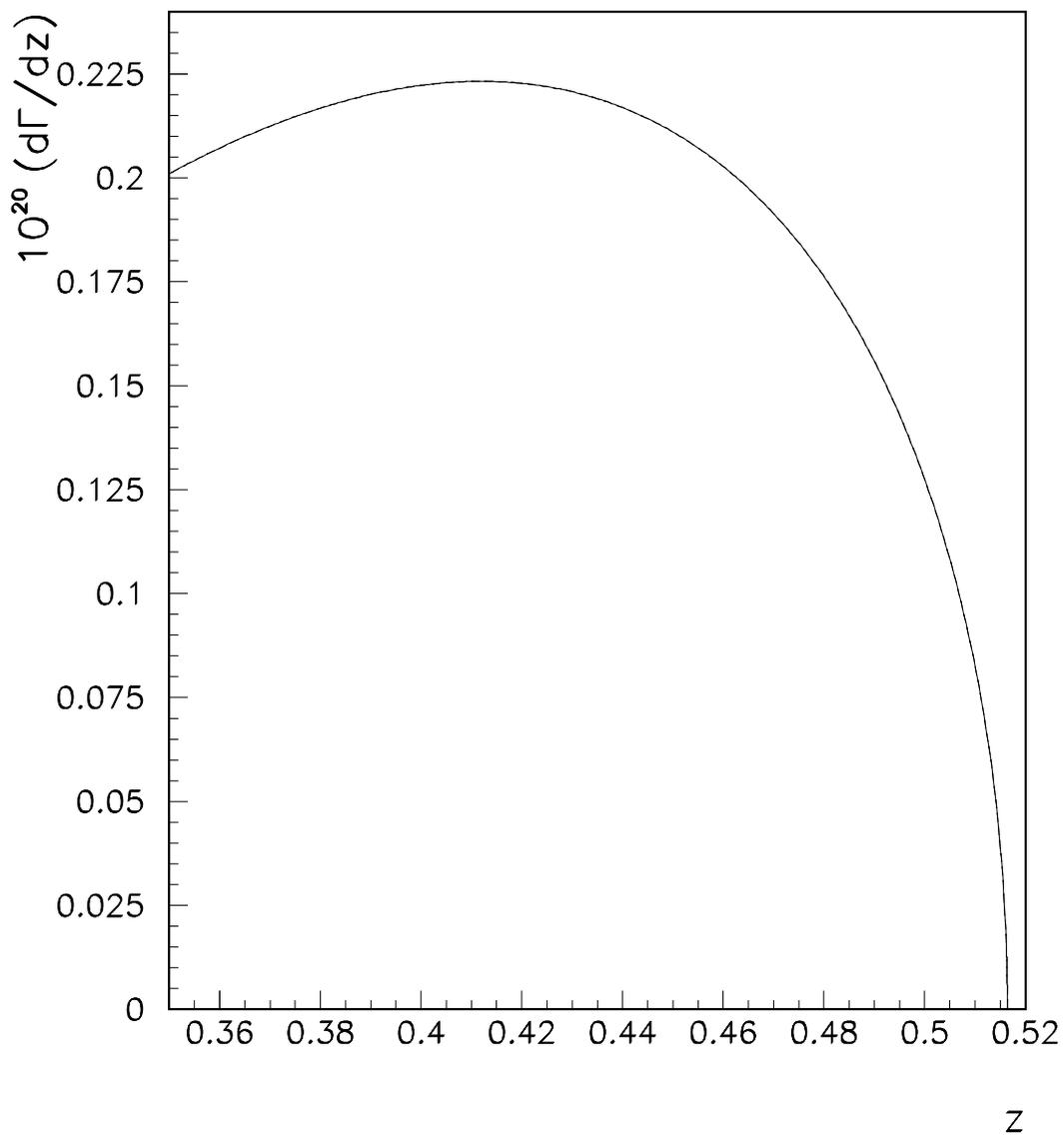}
\end{center}
\caption{Differential decay rate for $K^+\to\pi^+\mu^+\mu^-$ as a
function of $z$ with $0.35\le z\le (1-r_\pi)^2$. }
\end{figure}

\begin{figure}[t]
\begin{center}
\includegraphics[width=14cm,height=15cm]{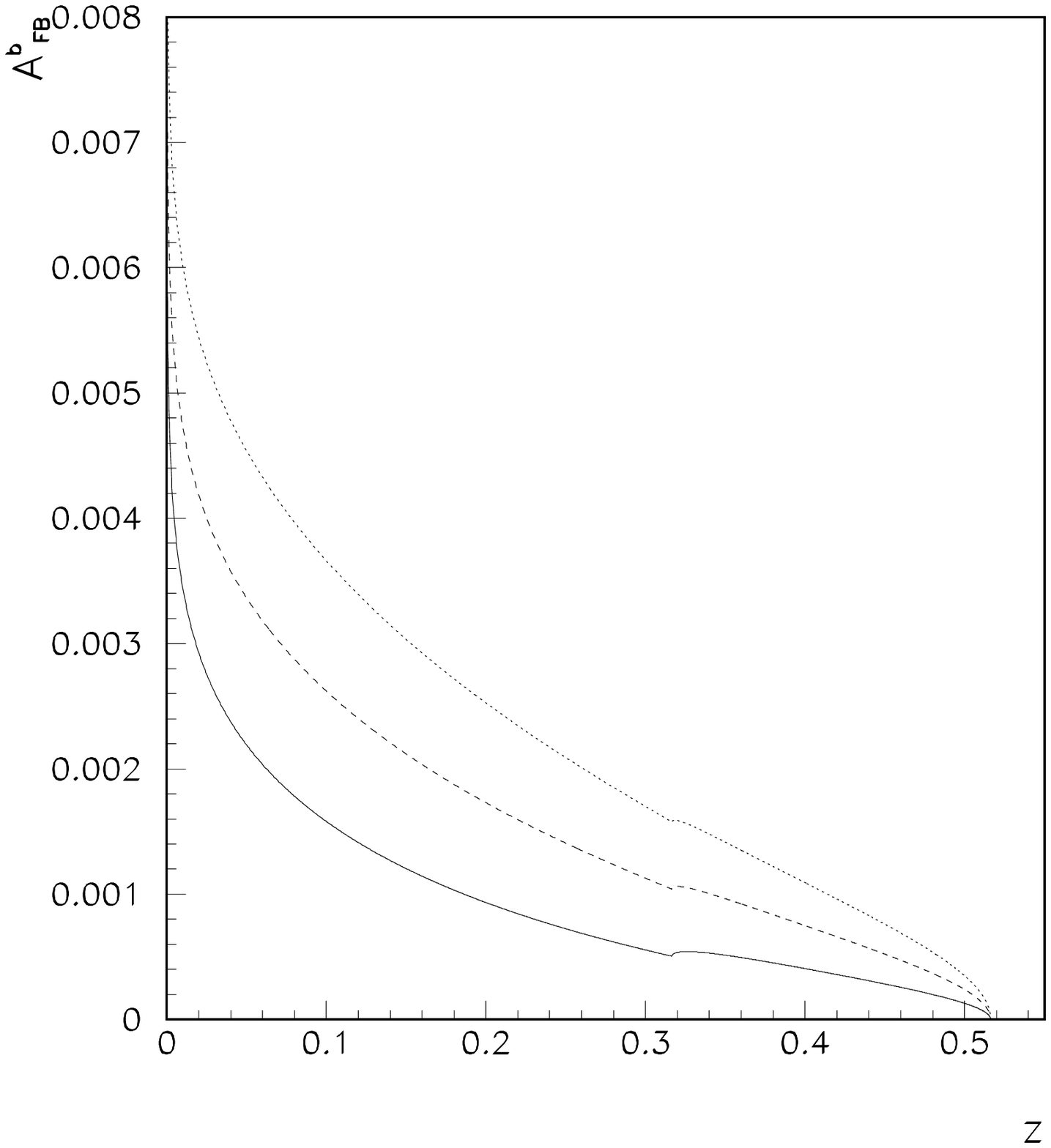}
\end{center}
\caption{Forward-backward asymmetry $A^b_{\rm FB}$ in decay
$K^+\to\pi^+e^+e^-$ as a function of $z$. The full line is for
$a=-1$, the dashed line for $a=0$, and the dotted line for $a=1$.
}
\end{figure}

\begin{figure}[t]
\begin{center}
\includegraphics[width=14cm,height=15cm]{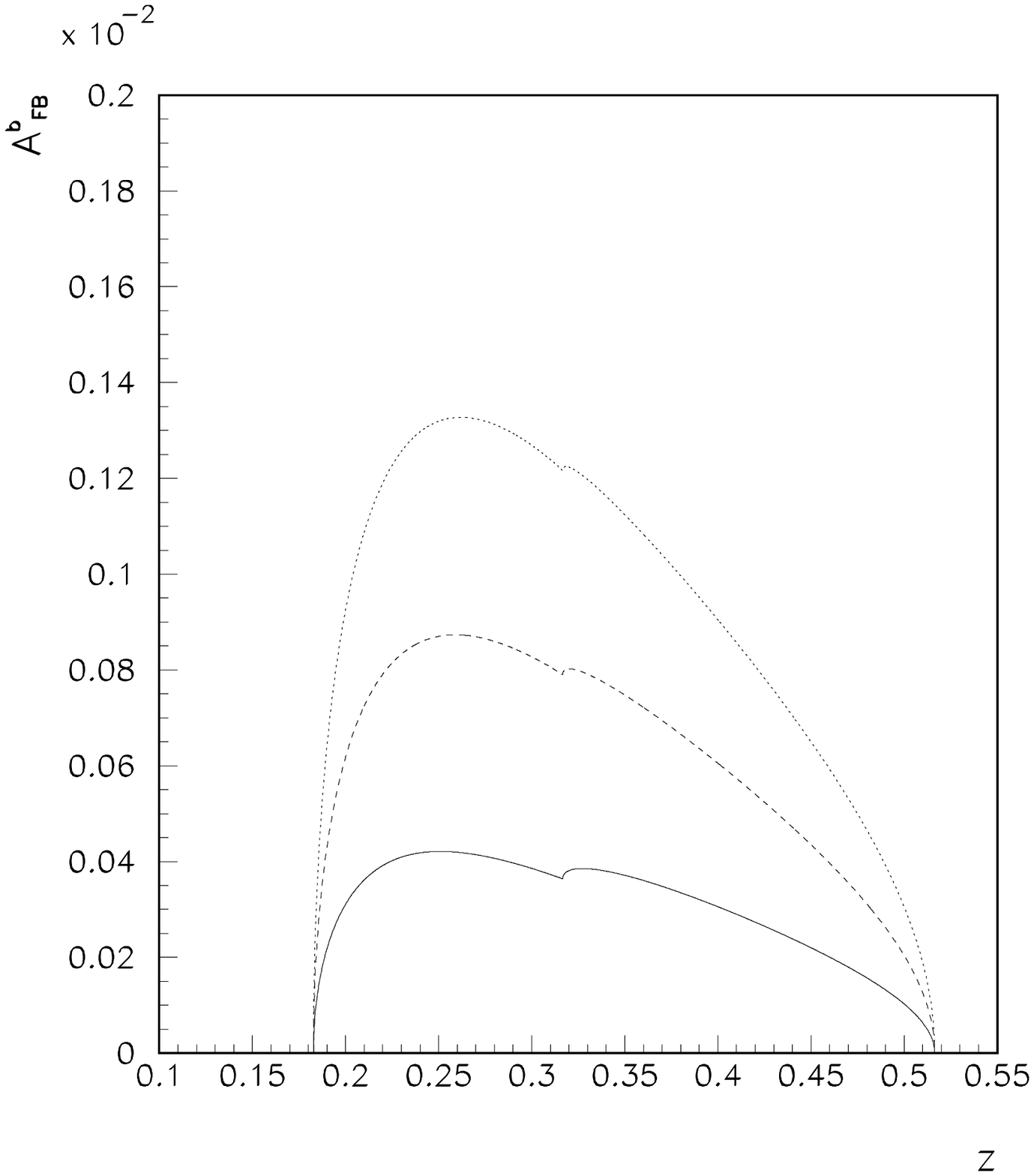}
\end{center}
\caption{Forward-backward asymmetry $A^b_{\rm FB}$ in decay
$K^+\to\pi^+\mu^+\mu^-$ as a function of $z$. The full line is for
$a=-1$, the dashed line for $a=0$, and the dotted line for $a=1$.
}
\end{figure}

In summary, we have studied the long-distance contribution via the
two-photon intermediate state to the forward-backward asymmetries
in decays $K^+\to\pi^+\ell^+\ell^-$. In order to estimate the
dispersive part of the
$K^+\to\pi^+\gamma^*\gamma^*\to\pi^+\ell^+\ell^-$ amplitude, a
phenomenological parameterization of the
$K^+\to\pi^+\gamma^*\gamma^*$ form factor has been used. Our
analysis shows that these asymmetries $A^a_{\rm FB}$ and $A^b_{\rm
FB}$ could be accessible to future experiments such as the CKM
experiment at Fermilab, where on the order of $10^5$ events for
these decays can be produced \cite{hyperCP02}. It is found that,
however, at present the theoretical uncertainty from the parameter
$a$ may obscure the standard model prediction to these quantities.
Therefore further study on the general parameterization of the
$K^+\to\pi^+ \gamma^*\gamma^*$ form factor both experimentally and
theoretically is needed to improve our understanding of the
forward-backward asymmetries in $K^+\to\pi^+\ell^+\ell^-$ decays.

\begin{center}{\bf ACKNOWLEDGMENTS}\end{center}
This work was supported in part by the NSF of China under Grant
No. 10275059.

\end{document}